\documentclass[%
 reprint,
 amsmath,amssymb,
 aps,
]{revtex4-2}

\usepackage[utf8]{inputenc}
\usepackage{graphicx}% Include figure files
\usepackage{dcolumn}% Align table columns on decimal point
\usepackage{bm}% bold math\usepackage{color,soul}
\usepackage{xcolor}
\usepackage{color,soul}
\usepackage{amssymb}
\usepackage{multirow}
\usepackage{float}
\usepackage[normalem]{ulem}
\begin{document}

\preprint{APS/123-QED}

\title{Extreme Cross-polarization Extinction Method for \\Quantification of Residual Crystal Anisotropy}% Force line breaks with \\

\author{Upasana Baishya}
\author{Nirmal K Viswanathan}%
 \email{nirmalsp@uohyd.ac.in}
\affiliation{%
 School of Physics, University of Hyderabad, Hyderabad, Telangana 500046, India
 }%

\date{\today}% It is always \today, today,
             %  but any date may be explicitly specified

\begin{abstract}
The high-extinction cross-polarization measurement of a paraxial laser beam results in a Hermite-Gaussian ($\text{HG}_{11}$)-like four-lobes with a dark-centered cross pattern. The azimuthally spin-separated mode pattern is understood to result from the spin-orbit interaction of light. The perfect dark region at the beam center, due to extreme cross-polarization extinction (EXE), is used here to accurately determine the on-axis residual polarization of the laser beam and, using it, the residual elliptical birefringence of optical crystals. By propagating the laser beam through uniaxial optical crystals with their optic axis cut either parallel or perpendicular to the beam propagation direction and performing the EXE measurement, we demonstrate a simple approach to precisely quantify the residual elliptical birefringence of crystals, which are otherwise assumed to be absent, to a high degree of extinction of 1 part in $10^{-8}$. The experimental results are interpreted using Jones-matrix-based calculations for a paraxial laser beam and upon its propagation through the optical crystals. The high sensitivity and high accuracy of the EXE technique make it well-suited for measuring and characterizing anisotropy and its dispersion of monolayers, ultrathin films, and 2D materials.
\end{abstract}

%\keywords{Suggested keywords}%Use showkeys class option if keyword
                              %display desired
\maketitle

%\tableofcontents

\section{\label{sec:I} Introduction}
Understanding the effects of light propagation through an anisotropic inhomogeneous medium, such as a transparent, birefringent, and optically active crystal, is of immense interest to physicists, chemists, biologists, crystallographers, and engineers. The anisotropy characteristics of such media have been theoretically modeled, experimentally measured, and used in various applications, assuming plane-wave beams and ideal crystals. However, the simultaneous presence of \textit{linear birefringence} (LB) and \textit{circular birefringence} (CB) (or optical activity) in such crystals makes the crystal \textit{elliptically birefringent} (EB) and the natural polarization modes elliptical. Propagating a linearly polarized light through such a crystal results in an elliptically polarized output beam with the ellipse axis rotated away from the input polarization plane. Understanding the propagation of light through such a generic, inhomogeneous, anisotropic medium is quite complicated due to the helical trajectory of the elliptically polarized modes \cite{malykin2016vl}. Thus, propagating a plane-wave beam ($\bar{k_0}$ ) through the crystal whose \textit{optical axis} (OA) is cut perpendicular to or parallel to it minimizes the effect of EB, simplifying the crystals to be understood as linear birefringent or circular birefringent, ignoring the effects due to CB and LB, respectively. Such crystals with ignorable weak \textit{elliptic birefringence} are relatively easy to model and understand under the assumption that the LB and CB effects are independent of each other \cite{ramachandran1961crystal,nye1985physical}. Several experimental techniques have been developed and demonstrated to measure the elliptical birefringence of such media under the assumption that the effects due to CB $\ll$ LB \cite{kobayashi1983new,chou1997effect,arteaga2009determination,martin2021chiroptical,yang2023simultaneous}. However, not ignoring the overall effects due to the weak EB results in an output beam whose state of polarization is elliptical and its plane of polarization is rotated with respect to the input linear polarized plane wave beam. The effects are characterized by the azimuth angle $\phi$ and the ellipticity angle $\epsilon$ of the eigen-polarization state and are experimentally measured to characterize crystal birefringence. Though weak, these effects are present in almost all commercial zero- and multi-order waveplates and rotators, but are typically ignored due to the inability of existing measurement techniques to quantify.

Instead of an ideal plane-wave beam, we consider here a paraxial (weakly-diverging) beam of light from a laser and its propagation through an anisotropic crystal with its OA oriented either parallel or perpendicular to the surface. A paraxial beam of light from a laser with a Gaussian intensity distribution and a finite angle of divergence has a spherical wavefront \cite{levy2019mathematics}. To satisfy the Maxwell equations and the transversality of the electromagnetic field, such a beam should have a nonzero longitudinal component ($E_z$) and a transverse cross-polarization component ($E_y$) in addition to the input field component ($E_x$), and the component field strengths are individually invariant under free-space propagation \cite{simon1987cross}. Passing such a beam through an ideal polarizer, crossed with respect to the principal polarization component of the beam, transmits only the Hermite-Gaussian ($HG_{11}$) like cross-polarization component \cite{fainman1984polarization,erikson1994polarization}. The vector superposition of the simultaneously present principal ($HG_{00}$) and cross polarization ($HG_{11}$) components in any cross section of the propagating beam shows that the polarization state of the laser radiation is spatially inhomogeneous \cite{sokolov2002polarization}. Under the assumption that the polarization inhomogeneity is low, the effect has been neglected, and a laser beam is approximated as a scalar wave-beam. However, a careful polarization analysis of a partially polarized laser beam transmitted through a quarter-wave plate (QWP) and a polarizer (P) shows that the scalar beam assumption of a typical laser system is incorrect. In addition, the phase difference between the propagating orthogonal polarized HG modes results in an elliptically polarized output beam with a transverse position-dependent ellipticity angle \cite{sokolov2002polarization}. Although these free-space propagation effects are very small, the transmission of the paraxial beam through a crystal amplifies them via the accumulation of the Pancharatnam-Berry type geometric phase.

Propagating through uniaxial crystals, the accumulation of the phase difference between orthogonally polarized \textit{ordinary} (o) and \textit{extraordinary} (e) ray-beams results in distance-dependent exchange of spin and orbital AM \cite{ciattoni2002paraxial} in the beam. This has led to the generation of optical vortex \cite{ciattoni2003circularly} and polarization-singular beams \cite{volyar2002vector} of light. Thus, a modern approach to understanding EB in optical crystals and its spatial variation is based on the \textit{spin-orbit interaction} (SOI) of light \cite{allen2003optical}. Although the spin and the (extrinsic) orbital angular momentum (AM) components of light, respectively, the circular polarization ($\sigma = \pm1$) and trajectory ($\bar{k}$) are separable in a plane wave beam \cite{allen2003optical} and upon its propagation through homogeneous – isotropic medium \cite{simpson1997mechanical}, they become coupled in inhomogeneous – anisotropic media. Optical fibers \cite{liberman1992spin}, space variant birefringent \cite{marrucci2006optical}, form-birefringent \cite{lerman2008generation} optical elements are some of the early examples of spin-orbit converters for a paraxial beam.

While the cross-polarization component of the laser is a $HG_{11}$ type four-lobe pattern in the linear basis, it is a Laguerre-Gaussian (LG) like vortex beam of charge $l=\pm2$ in the circular basis, with a perfect null at the beam center. Understood as due to SOI, the appearance of the $HG_{11}$ or the $LG_{20}$ beam at the focus or due to propagation through a crystal has been used to demonstrate the spin-Hall effect \cite{baishya2022dark} and spin-to-orbit conversion \cite{kumar2025spin}. The appearance of a perfect intensity null at the center of the beam provides the best contrast (relative to the rest of the beam intensity) for high-resolution measurements. The dark cross region has been used for the detection of nanoparticles \cite{hong2011background}, to measure the photonic spin-Hall effect induced by surface reflection \cite{baishya2022dark}, and to measure surface chirality \cite{baishya2023measurement}. These measurements demonstrate the importance of the perfect null intensity along the beam axis, enabled by the cross-polarization technique and its significant advantages.

\section{\label{sec:II} Working principle}
The principle behind the proposed technique and the experimental demonstration is to measure the cross-polarization characteristics of a partially polarized paraxial Gaussian laser beam and use it as a reference to quantify the residual EB of quartz and calcite crystals whose OA are either along or perpendicular to the beam propagation direction. The high spatial resolution (of the size of the CCD pixel $\sim 5\mu$m) and the sensitivity (up to $10^{-8}$, relative to parallel polarization and down to the electronic noise floor of the CCD camera) \cite{benelajla2021physical, steindl2023cross}, provide unprecedented precision for the method, termed \textit{the extreme cross-polarization extinction} (EXE) method. The EXE method is two-dimensional, which allows us to measure and quantify the polarization ellipticity ($\epsilon$) and the polarization ellipse orientation (azimuth angle) ($\phi$) with very high precision in the dark cross-region of the beam. Using the laser beam's cross-polarization characteristics, the residual EB of crystals with different orientations is measured. The polarization characteristics of the paraxial beam of light after passing through the crystal are simulated using the Jones matrix method, and the results match well with the experimental measurements. The measured and calculated crystal birefringence values are compared with those in the literature.
 
%%%%%%%%%%%%  Fig1  %%%%%%%%%%%%%%%%%
\begin{figure}[htbp]
\includegraphics[width=\columnwidth]{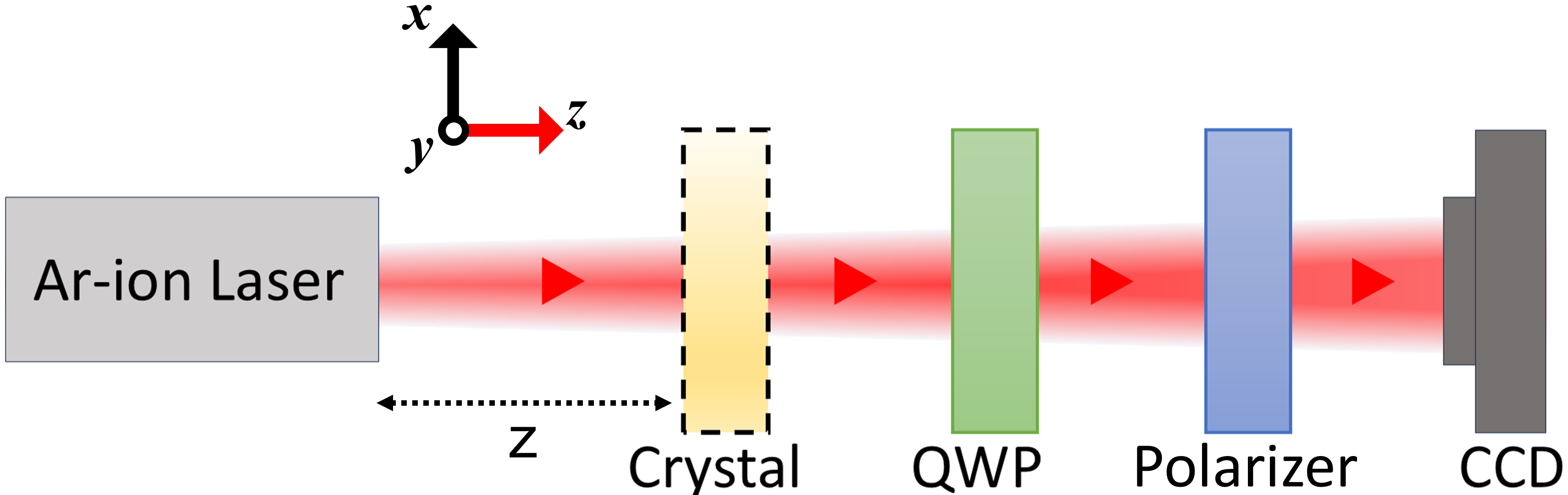}% Here is how to import EPS art
\caption{\label{fig1} Schematic of the optical setup used in the simulation and experiment. $Ar^+$ -laser of wavelength $\lambda$ = 514.5 nm, sample EB crystal, QWP- quarter-wave plate, P- polarizer, CCD- camera.}
\end{figure}
%%%%%%%%%%%%  Fig1  %%%%%%%%%%%%%%%%%

The accurate measurement of the polarization characteristics of the laser beam is the basis for the high-sensitivity quantification of the weak residual anisotropy of EB crystals. Our first task is thus to measure the polarization characteristics of the laser beam in detail and to establish the baseline settings for the EXE technique. The schematic of the optical system used is shown in Fig. \ref{fig1}. The paraxial Gaussian intensity beam from a $Ar^+$ laser ($\lambda$ = 514.5 nm) with a divergence angle of $0.036^\circ$ serves as the light source for the development of the EXE method. The Brewster cavity with a dispersion prism (for wavelength selection) produces a spatially partially polarized laser beam. A CCD camera is used to observe and measure the output beam intensity and its characteristics. To implement the EXE technique and measure the inhomogeneity in the state of elliptical polarization (SoEP) of the light beam, we use spatially resolved Stokes polarimetry and weak polarimetry with a quarter-wave plate (QWP) and a polarizer placed before the CCD. Subsequently, sample crystals are kept in the laser beam as shown in Fig. 1, to measure the modifications imparted on the beam's polarization characteristics to accurately quantify the residual polarization ellipticity $(\epsilon)$ and the ellipse orientation $(\phi)$ along the mean beam axis ($\langle \vec{k_0} \rangle$). Knowing both polarization parameters under EXE conditions allows us to extract high-sensitivity information on the residual EB properties of the crystal, as discussed.

\section{\label{sec:theory} Theory}

The polarization characteristics of the paraxial laser beam is modeled using the vectorial angular spectrum method of Richards and Wolf \cite{richards1959electromagnetic}. The spatial distribution of the electric field components in the focal region $\mathbf{E}_f(\rho, \psi, Z)$ is given by

    \begin{align} 
        \mathbf{E}_f & =  C_1 \begin{bmatrix}
        I_0 + I_2 e^{-2i\psi} \\
        -i I_0 + i I_2 e^{-2i\psi} 
        \end{bmatrix} 
        +
       C_2 \begin{bmatrix}
        I_0 + I_2 e^{2i\psi} \\
        i I_0 - i I_2 e^{2i\psi}
        \end{bmatrix} 
        \label{eq 1}
    \end{align}  

Where     \begin{align}
        I_0 & = \int_0^{\theta_{\text{max}}} A(\theta,Z) (1 + \cos\theta) J_0(k\:\rho\:sin\theta) \; d\theta \notag \\
        I_1 & = \int_0^{\theta_{\text{max}}} A(\theta,Z) \cos{\theta} J_1(k\:\rho\:sin\theta) \; d\theta \notag \\
        I_2 & = \int_0^{\theta_{\text{max}}} A(\theta,Z) (1 - \cos\theta) J_2(k\:\rho\:sin\theta) \; d\theta \notag\\
        A&(\theta ,Z) = \cfrac{i\:\pi}{\lambda}\: f\;E_\circ \; \sqrt{cos\theta}\:sin\theta\:e^{-ik\,Z\,cos\theta} \notag\\
        C_1 &= e^{i\beta} \cos\alpha \, \cfrac{1}{\sqrt{2}} \; \text{and} \; C_2 = e^{-i\beta} \sin\alpha \, \cfrac{1}{\sqrt{2}} \notag
    \end{align} 
Here, $\alpha \in [0,\cfrac{\pi}{2}]$ is the polarization ellipticity, $\beta \in [0,\pi)$ the azimuthal orientation of the polarization ellipse. Together, $\alpha$ and $\beta$ are used to characterize the SoEP of the laser beam. To model a TM-polarized input beam (linear polarization parallel to the x-axis), we set $\beta = 0^\circ$ and $\alpha = 45^\circ$. The polarization state of the input beam is then varied using the ellipticity angle $\alpha$ and the orientation angle $\beta$, to simulate experimental beam conditions with greater accuracy. Here, $Z$ is the axial distance from the laser output, with $Z=0$ corresponding to the laser output plane and $Z > 0$ indicating positions beyond, along the direction of propagation of the beam.

\section{\label{sec:laser characterisation} EXE technique to characterize laser beam}

\begin{figure}[htbp]
\includegraphics[width=\columnwidth]{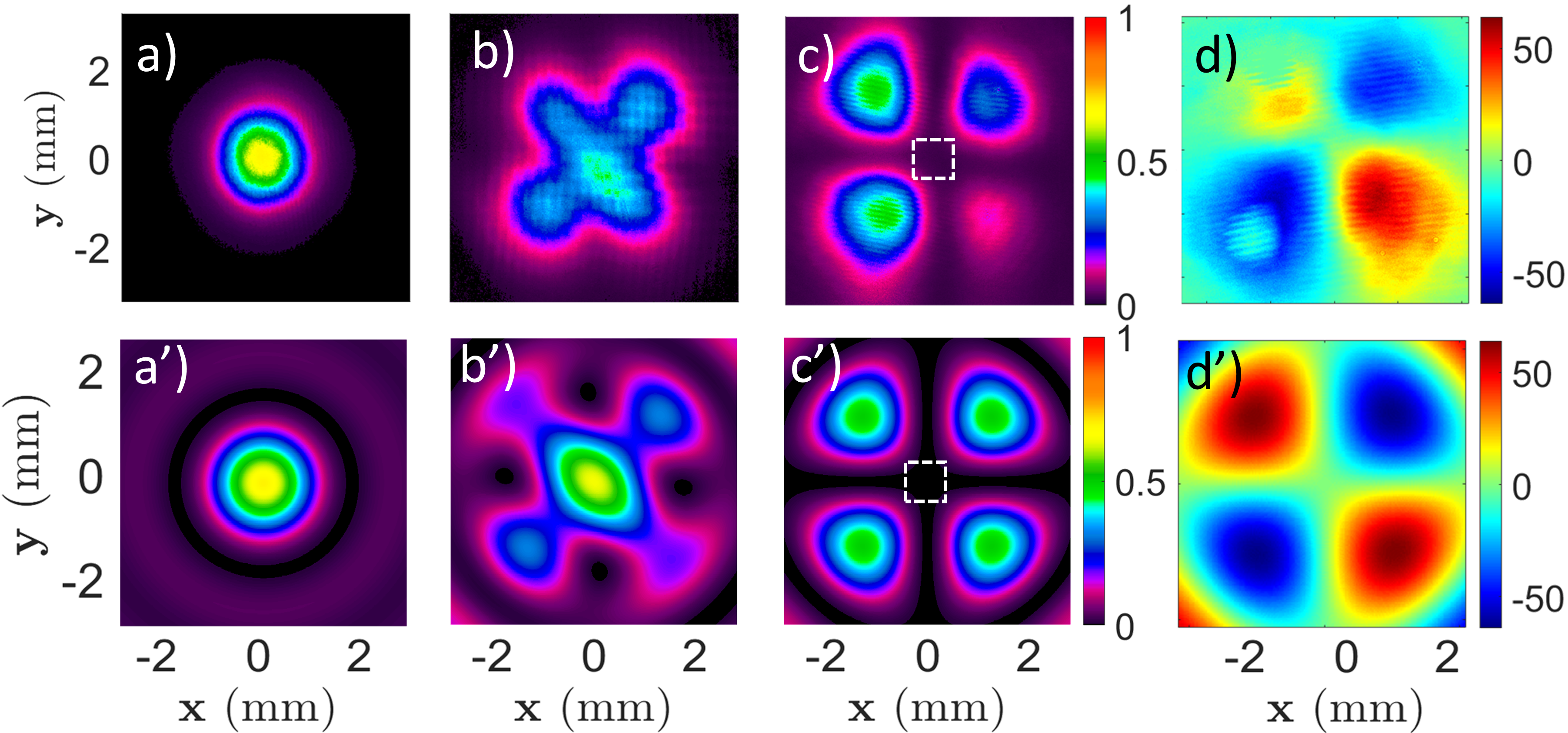}% Here is how to import EPS art
\caption{\label{fig2} Experimentally measured (a-b) and simulated (a’-b’) transverse intensity distribution of the paraxial laser beam for different orientation of the polarization elements. After passing through the optimized analyzer and QWP angles, the total intensity of the beam becomes a minimum, and the non-uniform cross-polarization intensity patterns of (b) and (b') become a four-lobe structure (c), (c'), due to the compensation of the non-uniform elliptical polarization of the laser beam. (d) and (d’) are the experimental and simulated $S_3$ Stokes parameter showing azimuthal spin variation.}
\end{figure}

We first investigate the polarization characteristics of the $Ar^+$ laser beam using only the polarizer and the CCD camera (Fig. \ref{fig1}). With the Glan-Thomson polarizer (GTH10M, Thorlabs, USA; with $1:10^6$ extinction ratio) mounted on a NanoRotator Stage (NR360S/M, Thorlabs, USA) set to $0^\circ$, the TM polarized beam intensity recorded by the CCD shows a Gaussian intensity profile (Fig. \ref{fig2}a), consistent with the expected fundamental mode of the laser. Rotating the polarizer to $90^\circ$ reveals a distorted beam structure that indicates non-zero polarization ellipticity in the beam cross-section with residual central intensity (Fig. \ref{fig2}b). A QWP (AQWP05M-630, Thorlabs, USA) mounted on a motorized rotation stage (K10CR1/M, Thorlabs, USA) is inserted before the polarizer and both elements are rotated simultaneously to minimize the total beam intensity, resulting in a clear four-lobe pattern and a dark cross pattern (Fig. \ref{fig2}c). The analyzer and QWP angles to achieve a perfect dark center in the cross pattern were $90.374^\circ$ and $-1.303^\circ$, respectively (Table \ref{table 1}). Although the unequal intensity of the four lobes is attributed to the non-uniform gain in the beam cross-section, the stability of the lobes confirms that the effect originates from the intrinsic polarization variation of the laser beam and its divergence \cite{fainman1984polarization, erikson1994polarization}.

The experimentally measured beam profiles are simulated considering the laser beam with a divergence angle of $0.036^\circ$ and a distance of $Z = 2$ m. Using the polarization parameters ($\alpha$, $\beta$), simulations were performed using the vectorial Debye-Richards–Wolf model described by Eq. \ref{eq 1}. The azimuthal orientation of the polarization ellipse $\beta$ is taken to be $0^\circ$, and with $C1 = \cfrac{1}{\sqrt{2}}\: \cos \alpha = a_1$ and $C2 = \cfrac{1}{\sqrt{2}}\: \sin \alpha = a_2$, Eq. \ref{eq 1} becomes
\begin{align} 
        \mathbf{E}_f & =  \begin{bmatrix}
        (a_2+a_1)(I_0 + I_2 \cos 2 \psi) + i (a_2 -a_1) I_2 \sin 2\psi  \\
        i (a_2-a_1)(I_0 + I_2 \cos 2 \psi) +  (a_2 + a_1) I_2 \sin 2\psi
        \end{bmatrix} 
        \label{eq 2}
    \end{align}  
The Jones matrix for the polarizer and the QWP is
\begin{align}
    J_{QWP} = & \cfrac{1}{\sqrt{2}} 
    \begin{bmatrix}
    (1-i) \cos 2 \theta_Q & -i\sin 2 \theta_Q \\ -i\sin 2 \theta_Q & (1+i) \cos 2 \theta_Q
    \end{bmatrix} \label{eq 3} \\ 
    J_P = & 
    \begin{bmatrix}
        \cos^2\theta_P & \cos\theta_P \sin\theta_P \\ \cos\theta_P \sin\theta_P & \sin^2\theta_P
    \end{bmatrix} \label{eq 4}
\end{align}

To obtain the simulated TM component, the beam (Eq. \ref{eq 2}) is passed through the polarizer oriented at $\theta_P = 0^\circ$. Using $\alpha =  44.626^\circ$, gives $(a_2-a_1) = 0.0007$ and $(a_2+a_1) = 0.999$ and Eq. \ref{eq 2} simplifies to
\begin{align} 
        \mathbf{E}_{fTM} & =  \begin{bmatrix}
        (0.999)(I_0 + I_2 \cos 2 \psi) + i (0.0007) I_2 \sin 2\psi  \\
        0
        \end{bmatrix} \notag
    \end{align}  
The simulated TM component (Fig. \ref{fig2}a') reproduces the Gaussian beam intensity distribution, since the dominant term of $|E_{fTM}|^2$ is $I_0$. 

The orthogonal cross-polarization component (Fig. \ref{fig2}b') obtained with $\alpha =  44.626^\circ$, and the polarizer angle $\theta_P = 90^\circ$ gives 
\begin{align} 
        \mathbf{E}_{fTE} & =  \begin{bmatrix}
        0 \\
        i (0.0007) (I_0 + I_2 \cos 2 \psi) +  (0.999) I_2 \sin 2\psi
        \end{bmatrix} \notag
    \end{align}
It matches the distorted four-lobe pattern, as the dominant term here is $I_2$, and $\sin 2 \psi$ gives alternate intensity maxima with $\psi \in [0, 2\pi]$, along with non-zero $I_0$. 

To obtain the proper four-lobe pattern, a QWP is introduced before the polarizer at $\theta_Q= 0^\circ$. After the polarizer at $\theta_p = 90^\circ$, Eq. \ref{eq 2} becomes
\begin{align}
    \mathbf{E}_{fQTE} &=
        \begin{bmatrix}
            0\\[3pt]
    \begin{aligned}
        &\frac{(i-1)}{\sqrt{2}} (0.0007)\bigl(I_0 + I_2 \cos 2\psi\bigr) \\
        &\quad + \frac{(1+i)}{\sqrt{2}} (0.999)\, I_2 \sin 2\psi
    \end{aligned}
    \end{bmatrix} \notag
\end{align}    
The QWP compensated output, $|E_{fQTE}|^2$ - (Fig. \ref{fig2} c'),  shows a symmetric four-lobe structure.

%%%%%%%%%%%%  Fig2  %%%%%%%%%%%%%%%%%
%%%%%%%%%%%%  Fig2  %%%%%%%%%%%%%%%%%

From the experimental measurements, we calculated the Stokes parameter $S_3$ (Fig. \ref{fig2} d) using weak polarimetry to quantify the spatial distribution of polarization ellipticity or spin components in the 4-lobe pattern \cite{baishya2022dark}. The two images of the output beam intensity corresponding to the polarizer angles of $90.374\pm  0.099^\circ$ and the fixed QWP angle of $-1.303^\circ$ are used to obtain the $S_3$ image shown in (Fig. \ref{fig2}d), which matches well with the simulated result (Fig. \ref{fig2}d'). Measurement of $S_3$ clearly shows that the 2 sets of diagonal lobes are orthogonal, elliptically polarized with a phase difference of $\pi$. In Fig.\ref{fig2} (c) and (c') a white dotted line square of size 20 x 20 pixels ($93 \mu m \times 93 \mu m$) is marked at the center of the dark cross pattern, where all subsequent intensity measurements and simulations are carried out. This region corresponds to an angular divergence of $0.0013^\circ$ around the beam axis, $k_0$. The fields are taken to be transverse $(E_x, E_y)$ and the role of the axial $(E_z)$ component of the field is neglected in all our calculations and measurements in the marked region.

%%%%%%%%%%%%  Fig3  %%%%%%%%%%%%%%%%%
\begin{figure}[htbp]
\includegraphics[width=\columnwidth]{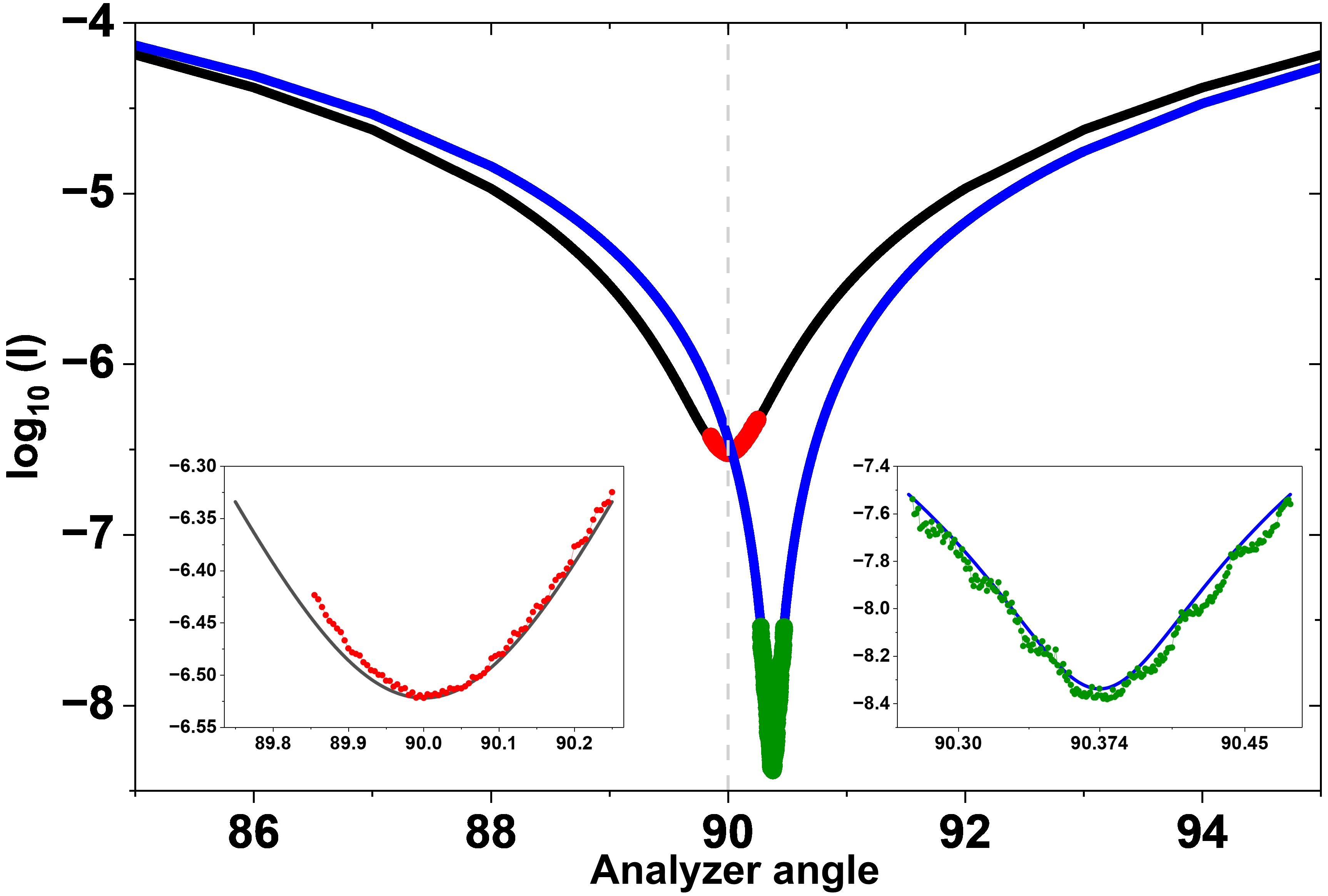}% Here is how to import EPS art
\caption{\label{fig3} Experimentally measured and simulated logarithmic intensity as a function of analyzer angle. The conventional polarization analysis, by rotating the polarizer around the cross-position of $90^\circ$ is shown by red circles (experiment) and a black continuous curve (simulation), with the left inset providing a magnified view. The extreme cross-polarization extinction (EXE) method, using optimized orientations of the QWP and polarizer, is shown by green circles (experiment) and a blue continuous curve (simulation), with the inset on the right providing a magnified view. The average intensity values are calculated from the $20 \times 20$ pixel region ($93,\mu\text{m} \times 93,\mu\text{m}$) centered on the beam axis. The EXE curve exhibits a sharper minimum, enabling more precise determination of the angles and, hence, the polarization ellipticity and ellipse orientation.}
\end{figure}
%%%%%%%%%%%%  Fig3  %%%%%%%%%%%%%%%%%

The dynamic range of the CCD-measured intensity in the marked region varies by almost eight orders of magnitude between the parallel and orthogonal polarizer orientations. To ensure the effectiveness of the EXE method, we calibrate the CCD using neutral density filters (NDFs) with an OD of 5.56 in parallel orientation, remove all of them, and operate the CCD in the highest sensitivity regime for orthogonal polarizer orientation. The logarithmic plot of the measured intensity as a function of the polarizer angle is shown in Fig. 3 without and with the oriented QWP. The intensity of the beam considered here corresponds to the white dotted-line square region shown in Fig. 2 (c) and (c'). For the parallel-polarizer orientation, the intensity at the beam center for cross-polarization decreases to the detector's electronic noise level. From the plots in Fig. 3 it is seen that the difference in the intensity versus the polarizer angle corresponding to the minimum intensity position without and with the QWP clearly shows that the minimum measurable intensity in the beam center is ~3 orders of magnitude lower, and the polarizer angle where the minimum intensity is measured is slightly different. Replacing the CCD with a point detector with a lower noise floor will allow one to measure the QWP and polarizer angles with much higher accuracy and sensitivity. The angle of the polarization elements corresponding to the minimum intensity in the marked region is then used to extract the residual polarization ellipticity and the ellipse orientation of the light beam. This demonstrates the EXE method's ability to measure very small SoEP values of the beam from the CCD's lowest measurable intensity level.

The EXE technique involves extracting the intensity value from the dark region of a $20 \times 20$ pixel area centered on the beam, corresponding to a beam divergence angle of $0.0013^\circ$. For further calculations, we consider this region of the beam to be the beam axis $\langle k_0 \rangle$. From the QWP and polarizer angles corresponding to the null intensity, we reconstruct the SoEP of the beam at the beam center. The Jones vector of the beam is then determined from the QWP–polarizer angles in the null-intensity reconstruction method. For a given QWP angle $\theta_Q$ and analyzer angle $\theta_P$, a null in transmitted intensity implies that the input state $J_{in}$ satisfies
\begin{equation} \label{eq 5}
    P^\dagger (\theta_P) \: J_{qwp}(\theta_Q) \: J_{in} = 0 
\end{equation}
where P is the transmission axis of the polarizer and $J_{QWP}(\theta_Q)$ is the Jones matrix of the rotated quarter-wave plate. The input polarization, therefore, lies in the null-space of this linear mapping. Solving the homogeneous equation yields the normalized Jones vector (unique up to global phase). A consistency check is performed for both possible retardance signs of the QWP, and the physically correct one is selected as the solution that reproduces the experimentally observed null. The amplitude (a) and phase ($\Delta$) are calculated using the MATLAB program, and the Jones vector of the laser beam is determined using
\begin{align}  
    \begin{bmatrix}
        E_x \\ E_y
    \end{bmatrix}& = \begin{bmatrix}
        \cos a \\ \sin a \; e^{i \Delta}
    \end{bmatrix} \label{eq 6}
\end{align} 

From the field components and the phase difference between them, the Stokes parameters are calculated using
\begin{equation}
\begin{aligned}
S_0 = |E_x|^2 + |E_y|^2  \qquad
&S_1 = |E_x|^2 - |E_y|^2\\
S_2 = 2\,\Re\:\big(E_x E_y^{*}\big)  \qquad
&S_3 = 2\,\Im\:\big(E_x E_y^{*}\big)
\end{aligned}
\end{equation} \label{eq 7}

Using Stokes parameters $S_1, S_2, S_3$, the ellipticity and ellipse orientation of the polarization ellipse in the center of the beam are calculated using Eq (\ref{eq 8}) and (\ref{eq 9}) and the values are given in Table \ref{table 2}.

\begin{align}
\text{Ellipticity, }  \epsilon = &  \frac{1}{2} \sin^{-1} S_3 \label{eq 8}\\
\text{Ellipse orientation, } \phi = & \frac{1}{2} \tan^{-1} \cfrac{S_2}{S_1} \label{eq 9}
\end{align}

\section{\label{sec:Anisotropic material} Residual crystal anisotropy}
We now extend the application of the EXE method to extract the residual elliptical birefringence of different anisotropic crystals. Our samples are quartz and calcite, which are, respectively, positive ($n_e > n_o$) and negative ($n_e < n_o$) uniaxial crystals. The orientation of the OA with respect to the crystal surface or $\langle k_0 \rangle$ critically determines the birefringence-induced polarization transformation of the laser beam. We consider two representative cases: (i) OA parallel to the surface and (ii) OA perpendicular to the surface. In general, it is assumed that a quartz crystal cut with its OA perpendicular to the surface (and parallel to the beam axis) behaves as an optical rotator, rotating the plane of linearly polarized light propagating through it without introducing ellipticity. Although such an optical element is predominantly circular birefringent, high-sensitivity EXE measurements reveal that the output beam shows weak ellipticity with a rotated plane of polarization due to the simultaneous presence of weak linear birefringence, making it an EB crystal, even along the OA. Independently, the quartz and calcite crystals, cut with their OA parallel to the surface (and perpendicular to the beam axis), are shown to behave as elliptical polarization rotators due to weak elliptical birefringence rather than just linear birefringent systems. The presence of orders of magnitude weak residual linear or circular birefringence, respectively, in the above two cases, makes the crystals EB, in general. However, the measurement of orders of magnitude weak birefringence effects, in the presence of the dominant birefringence of the crystal, is quite challenging and is rarely considered to be measurable.

To enable measurement and calculation, we first outline the theoretical framework for a paraxial Gaussian beam transmitted through the anisotropic medium and use the EXE protocol to characterize weak crystal anisotropy. To model the propagation of light through these anisotropic crystals, we use the Jones matrix for the elliptical retarder, which accounts for the vectorial nature of the electric field and the anisotropic dielectric response, including residual effects. Considering the crystals as an elliptical retarder, with orthogonal eigenpolarizations $\chi_{ef}$, and $\chi_{es}$ (with $\chi_{ef} \:\chi_{es}^* = -1 $) and eigenvalues $e^{i \delta /2}$, and $e^{-i \delta /2}$, the transmission matrix is written as \cite{azzam1978ellipsometry} 
\begin{equation} \label{eq 10}
    T_{ER} = T_0 \begin{bmatrix}
        e^{i \frac{\delta}{2}} + \chi_{ef} \: \chi_{ef}^* \: e^{-i \frac{\delta}{2}} & 2i  \chi_{ef} \sin \frac{\delta}{2} \\
        2i  \chi_{ef} \sin \frac{\delta}{2} & e^{-i \frac{\delta}{2}} + \chi_{ef} \: \chi_{ef}^* \: e^{i \frac{\delta}{2}}
    \end{bmatrix}
\end{equation}
Where $T_0 = (1 + \chi_{ef} \: \chi_{ef}^*)^{-1}$ and $ \chi_{ef} = \Bigl(\cfrac{\tan \phi + i \tan \epsilon}{1- i  \tan \phi \tan \epsilon} \Bigl)$ with $\phi$, $\epsilon$ are the azimuth and ellipticity angles of $\chi_{ef}$ respectively.

\subsection{OA parallel to the surface: Linear retarder}

\begin{figure}[htbp]
\includegraphics[width=\columnwidth]{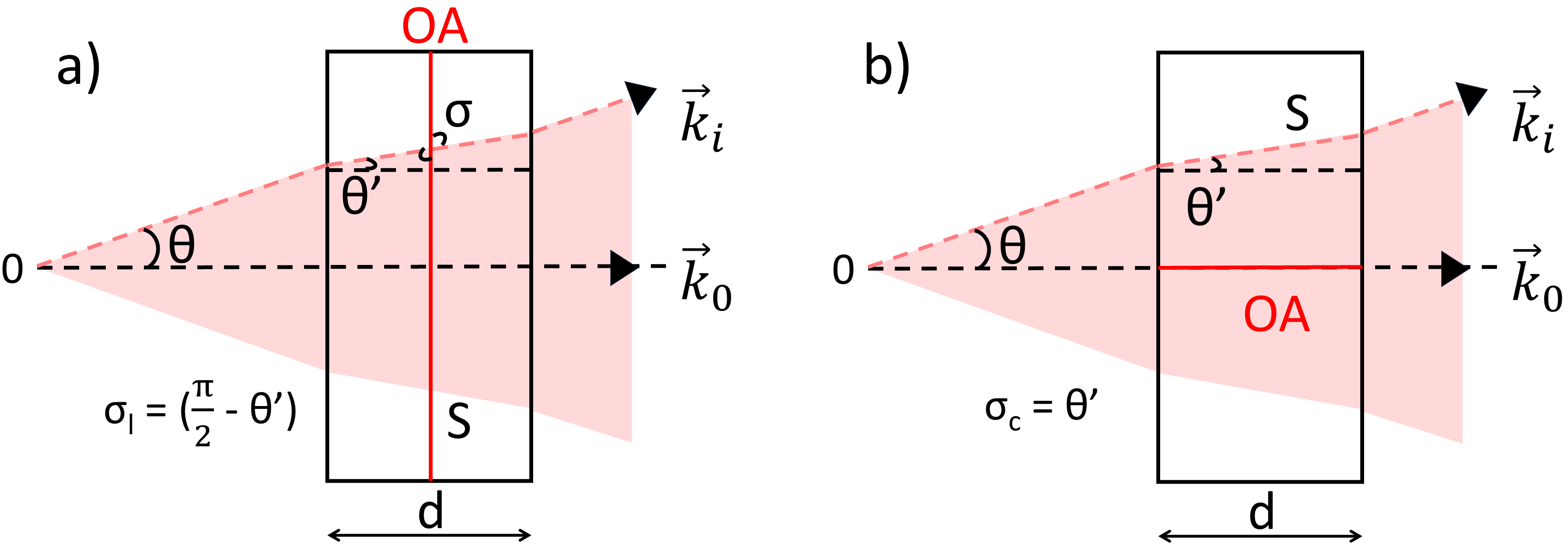}% Here is how to import EPS art
\caption{\label{fig4}  Schematic diagram of a paraxial beam of light from a laser propagating through a uniaxial crystal. (a) The optic axis (OA) of the crystal is parallel to the surface and perpendicular to $\textbf{k}_0$, 'd' is the physical thickness of the crystal. The off-axis wavevector $\textbf{k}_i$ makes an angle $\sigma_l=(\cfrac{\pi}{2}-\theta')$ with the OA. (b) The OA of the crystal is perpendicular to the crystal surface and parallel to $\textbf{k}_0$. The off-axis wavevector $\textbf{k}_i$ makes an angle $\sigma_c = \theta'$ with the OA.}
\end{figure}

Figure \ref{fig4} (a) shows the schematic for the linear retarder case where the orientation of the optic axis (OA) is parallel to the crystal surface. The coordinate system is defined such that the z-axis is the direction of beam propagation, and the OA lies in the x–y plane. The incident paraxial beam propagates through the crystal with a divergence angle of $\theta$. The angle the wavevector makes after entering the crystal is $\theta'$. The angle between the axial component of the wavevector $\textbf{k}_0$ and the OA of the crystal is $\frac{\pi}{2}$, propagating through the crystal, the off-axis wavevector $\textbf{k}_i$ makes an angle $(\sigma =\frac{\pi}{2}-\theta')$ with the OA. With $d$ as the thickness of the crystal and S the light path through the crystal, using Snell’s law, we get, \begin{equation*}
    \theta' = \arcsin \Big(\cfrac{\sin \theta}{n_o}\Big),  \hspace{0.5cm} S = \cfrac{d}{\cos \theta'}
\end{equation*} 

Except for $\textbf{k}_0$, the conical divergent beam propagating through the crystal accumulates radial phase differences due to $n_0$ and $n_e$ (dependent on $\theta'$), as the beam propagates through the crystal. This makes the output beam, after propagating through the crystal, elliptically polarized with spatially varying ellipticity and ellipse orientation. For a linear retarder, ellipticity, $\epsilon =0$, $\chi_{ef} = \tan \phi$.  Using this in Eq. \ref{eq 10}, the transmission matrix for a linear retarder becomes
\begin{align} \label{eq 11}
    T_{LR}  = & \begin{bmatrix}
         e^{i \frac{\delta}{2}} \cos^2 \phi + e^{-i \frac{\delta}{2}} \sin^2 \phi & 2i \sin\phi \cos \phi \sin \frac{\delta}{2} \\
        2i \sin\phi \cos \phi \sin \frac{\delta}{2} & e^{i \frac{\delta}{2}} \sin^2 \phi + e^{-i \frac{\delta}{2}} \cos^2 \phi
    \end{bmatrix} 
\end{align}
     The linear birefringence LB \cite{born2013principles} is given by
\begin{align}     
    & \delta(\theta')_{LB} = \cfrac{2 \pi}{\lambda} [n_e(\theta')-n_o] \: \cfrac{d}{\cos \theta'} \label{eq 12} \\
   & \text{and } \hspace{0.4 cm } \cfrac{1}{n_e^2(\theta')} = \cfrac{\sin^2 \theta'}{n_e^2} + \cfrac{\cos^2 \theta'}{n_o^2} 
\end{align} \label{eq 13}
For the central wave-vector $\textbf{k}_0$,
\begin{equation}
    \delta_{LB} = \cfrac{2 \pi}{\lambda} [n_e-n_o] \: d  \label{eq 14}
\end{equation}

%%%%%%%%%%%%  Fig4  %%%%%%%%%%%%%%%%%
%%%%%%%%%%%%  Fig4  %%%%%%%%%%%%%%%%%

\subsection{OA perpendicular to the surface: Circular retarder}

Figure \ref{fig4} (b) shows the schematic of the case where the OA is perpendicular to the crystal surface and along the propagation direction, $\textbf{k}_0$. The central wavevector $\textbf{k}_0$, is normally incident, forming an angle $0^\circ$, with the OA, while an off-axis wavevector  $\textbf{k}_i$, makes an angle $\theta'$ with the OA. In this configuration, the light beam propagating along the OA, and the polarization eigenstates are circular. The phase difference between the orthogonal circular eigenstates makes the component a polarization rotator, causing the linearly polarized light propagating along the OA to rotate the plane of polarization. However, the paraxial beam divergence introduces small angular deviation, leading to accumulation of spatially dependent elliptical birefringence as the right- and left-circularly polarized components accumulate unequal phases depending on the local incidence angle. For the circular retarder (or optical rotator), the ellipticity $\epsilon = 1$, $\phi = 0^\circ$ and $\chi_{ef} = \pm i$, and the corresponding transmission matrix is - 
\begin{align}
    & T_{CR} = \begin{bmatrix}
        \cos \frac{\delta}{2} & \pm \sin \frac{\delta}{2} \\
       \mp \sin \frac{\delta}{2} & \cos \frac{\delta}{2}
    \end{bmatrix}  \label{eq 15}
\end{align}
The circular birefringence, CB is given by \cite{arteaga2009determination},
\begin{align}
 \delta(\theta')_{CB} & = \cfrac{2 \pi}{\lambda} (n_L - n_R) \cfrac{d}{\cos \theta'} \label{eq 16}\\
 \delta(\theta')_{CB} & = \cfrac{2 \pi}{\lambda} \Big(\cfrac{g_{11} \sin^2 \theta ' + g_{33} \cos^2 \theta'}{\sqrt{n_o n_e}} \Big) \cfrac{d}{\cos \theta'} \label{eq 17}
\end{align} 
Here, $g_{11}$ and $g_{33}$ are the components of the crystal gyration tensor. For quartz,  $|g_{11}| = 6.1 \times 10^{-5}$ and $|g_{33}| = 12.81 \times 10^{-5}$ for $\lambda = 510 nm$ \cite{arteaga2009determination}.
For the central wave-vector, $\textbf{k}_0$ 
\begin{equation} \label{eq 18}
     \delta_{CB} = \cfrac{2 \pi}{\lambda} \cfrac{ g_{33} }{\sqrt{n_o n_e}} d
\end{equation}

\section{Calculation of Birefringence from the Transmitted Field} \label{section IV} 
We now describe the numerical procedure used to determine the elliptical birefringence due to the propagation of a paraxial beam of light through a uniaxial crystal. This is done from the measured transmitted optical fields. Consider an elliptically polarized input beam given by
\begin{equation}  \label{eq 19}
    \begin{bmatrix}
        E^x_{in} \\ E^y_{in}
    \end{bmatrix} = \begin{bmatrix}
        \cos a \\ \sin a \; e^{i \Delta}
    \end{bmatrix}
\end{equation}
For linear retarder, described by Eq. \ref{eq 11}, the transmitted output light field with the OA azimuth $\phi = 0$ is
\begin{align}
    \begin{bmatrix}
         E^x_{out} \\ E^y_{out}
    \end{bmatrix} & = \begin{bmatrix}
        e^{i \frac{\delta}{2}} & 0 \\ 0 & e^{-i \frac{\delta}{2}}
    \end{bmatrix} \begin{bmatrix}
        E^x_{in} \\ E^y_{in}
    \end{bmatrix}   \label{Eq 20}
\end{align} 
Solving this (Appendix \ref{appendixA}), gives the linear birefringence as 
\begin{equation}
    \delta_{LB} = \: i \: ln(E_L) + 2 \pi m \label{eq 21}   
\end{equation}
Where, $E_L = \cfrac{ E^x_{in} \; E^y_{out}}{E^y_{in} \; E^x_{out}}$ is a complex number.

For circular retarder, described by Eq. \ref{eq 15}, the transmitted output light field is 
\begin{equation} \label{eq 22}
        \begin{bmatrix}
         E^x_{out} \\ E^y_{out}
    \end{bmatrix} =   \begin{bmatrix}
        \cos \frac{\delta}{2} & - \sin \frac{\delta}{2} \\
       + \sin \frac{\delta}{2} & \cos \frac{\delta}{2}
    \end{bmatrix} \begin{bmatrix} E^x_{in} \\ E^y_{in} \end{bmatrix} 
\end{equation}
Solving this expression (Appendix \ref{appendixB}), gives the circular birefringence of the crystal as
\begin{equation}
    \delta_{CB} = \pm \cos^{-1} \Biggl( \frac{E_{C1}}{\sqrt{1+4  E^2_{C2}}} \Biggl)  - \tan^{-1} (2 E_{C2})  \label{eq 23}  
\end{equation}
with $E_{C1} = \biggl( \cfrac{{E^x_{out}}^2  - {E^y_{out}}^2}{{E^x_{in}}^2  - {E^y_{in}}^2} \biggl) \: ; \: E_{C2} = \biggl(\cfrac{ E^x_{in}  E^y_{in}}{{E^x_{in}}^2  - {E^y_{in}}^2}\biggl) $

\section{EXE technique in crystal transmission}
Next, we propagate the laser beam through different uniaxial crystals, as an application of the EXE method to extract weak residual crystal anisotropy. We consider uniaxial calcite and quartz crystals in which the OA is oriented perpendicular to the surface and right- (R) and left- (L)-handed quartz crystals and a calcite crystal. From Eqs. (\ref{eq 11}, \ref{eq 15}), it is seen that the sample thickness $d$ directly governs the effective birefringence and, hence, the polarization evolution within the crystal, resulting in markedly different polarization characteristics of the output beam.

\begin{figure}[htbp]
\includegraphics[width=\columnwidth]{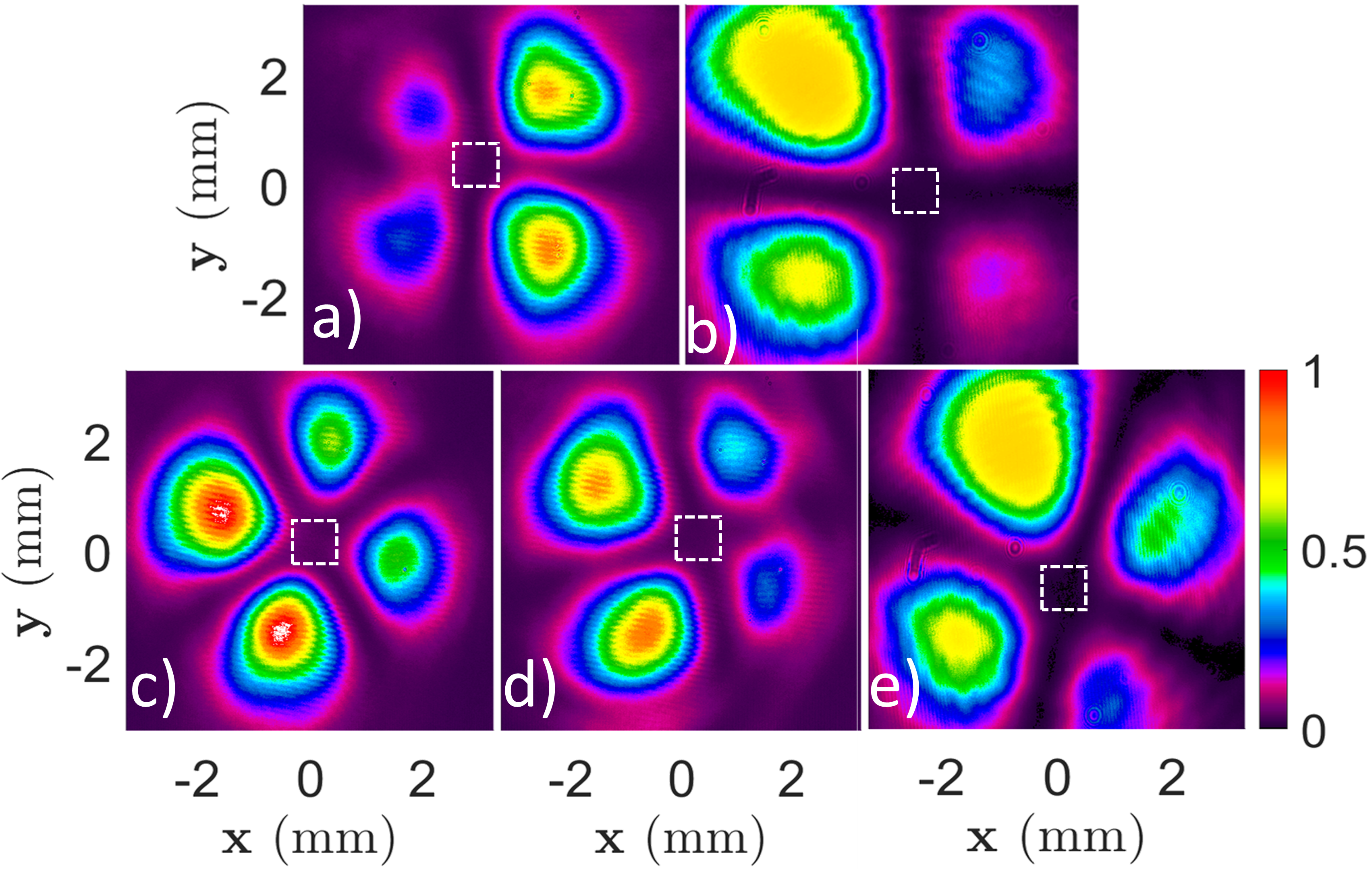}% Here is how to import EPS art
\caption{\label{fig5} Experimentally measured images of the four-lobes with dark centered cross-pattern under EXE conditions for (a) quartz, $Q_{||}$, and (b) calcite, $C_{||}$, with the OA parallel to the crystal surface; and for different thickness quartz crystals with the OA perpendicular to the surface: (c) $Q_{\perp1}$, (d) $Q_{\perp2}$, and (e) $Q_{\perp3}$.}
\end{figure}

As shown in Fig. \ref{fig1}, a QWP and an analyzer (polarizer) are placed after the crystal and in front of the CCD. With the crystal kept normal to the beam, the QWP - P are rotated sequentially to produce a four-lobe pattern with a dark-centered cross pattern in the output beam. From the QWP-P angle, we calculate the residual elliptical birefringence of the crystal along the axial direction, corresponding to the dark-beam center. We first used the quartz crystal, $Q_\parallel$ (SIO-1a, OPTOGAMA, Lithuania), and then the calcite crystal, $C_\parallel$ (CACO-1a, OPTOGAMA, Lithuania), both with their OA parallel to the crystal surface. The crystals were placed in the laser beam with the azimuthal angle of OA $\phi = 0^\circ$. Ideally, neither crystal is expected to introduce any additional polarization ellipticity to the transmitted beam, as they are fabricated to work as multi-order half-wave plates. However, for the paraxial beam from the laser propagating through the crystals, the angle of incidence, the accumulated phase difference, and, hence, the EB in the output beam exhibit spatial variations. Due to varying angles of incidence of the weakly divergent laser beam, different regions of the wavefront experience different amounts of EB, including along the propagation axis $\textbf{k}_0$. It is important to note here that for a known input laser beam parameters, the measured elliptical birefringence is attributed predominantly to the residual crystal characteristics, beyond what is expected of an ideal behavior. This results in a complex output beam with spatially non-uniform elliptical polarization. The QWP and analyzer angles are optimized to obtain a dark-centered cross pattern and four-lobe pattern, as shown in Figs. \ref{fig5} (a) and (b). The QWP and analyzer angles $(\theta_Q, \theta_A)$ to achieve the output beam pattern are listed in Table \ref{table 1}. We use these angles to quantify the polarization ellipticity, ellipse orientation, and hence the residual elliptical birefringence of the crystals (Table \ref{table 2}). The values obtained are compared with the literature and manufacturer-specified values. Since we are interested here only in the extraction of the axial elliptical birefringence of the crystals, we focus only on the dark-centered intersection of the cross-pattern and ignore their spatial variation, the slight non-uniformity and asymmetry in the four-lobe intensity pattern.

%%%%%%%%%%%%  Fig5  %%%%%%%%%%%%%%%%%
%%%%%%%%%%%%  Fig5  %%%%%%%%%%%%%%%%%

Next, we study two right-handed quartz samples $Q_\perp1 $ (SIO-1b, OPTOGAMA, Lithuania), $Q_\perp2 $ (300EXR, Rudolph Instruments, USA) and one left-handed quartz sample $Q_\perp 3$ (300EXL, Rudolph Instruments, USA) of different thickness, with their OA perpendicular to the surface of the crystal, along the beam propagation direction. Following the EXE procedure, we rotate the QWP and the analyzer sequentially, nullify the residual polarization ellipticity at the beam center, and obtain the angles at which the four-lobe structure with a dark-centered cross pattern is observed. The experimentally measured patterns are shown in Fig. \ref{fig5} (c), (d) and (e). Compared with the previous case of crystals with their OA parallel to the surface, the four lobes are rotated by different angles as a result of the crystal's optical activity. As these quartz crystals have different thicknesses and therefore rotate the polarization of the beam by different angles, the experimentally obtained angles $(\theta_Q, \theta_A)$ to obtain the patterns shown in  Fig. \ref{fig5} (c), (d) and (e) are also different and given in Table \ref{table 1}. Like in the previous case, we ignore the slight non-uniformity and asymmetry in the four-lobe pattern and focus only on the dark-centered intersection of the cross-pattern. The circular birefringence of a quartz rotator, cut with its OA perpendicular to the surface and kept along the beam axis ($\textbf{k}_0$), rotates the linear plane-polarized beam of light by a fixed amount that depends on its thickness. In our experiments, however, we use a paraxial beam of light rather than a plane wave propagating through the crystal. As a result, different parts of the paraxial laser beam with different amounts of SoEP (as was demonstrated in Sect. IV) pick up different amounts of polarization ellipticity while propagating through the crystals. This can be seen in Figs. \ref{fig5} (c), (d), and (e), where we notice that the off-axis variations in the four-lobe pattern are quite different due to the spatially varying coupling between the laser polarization and the crystal’s elliptical birefringence. We do not address these complex effects here and reserve them for future investigations.
 
\begin{figure}[htbp]
\includegraphics[width=\columnwidth]{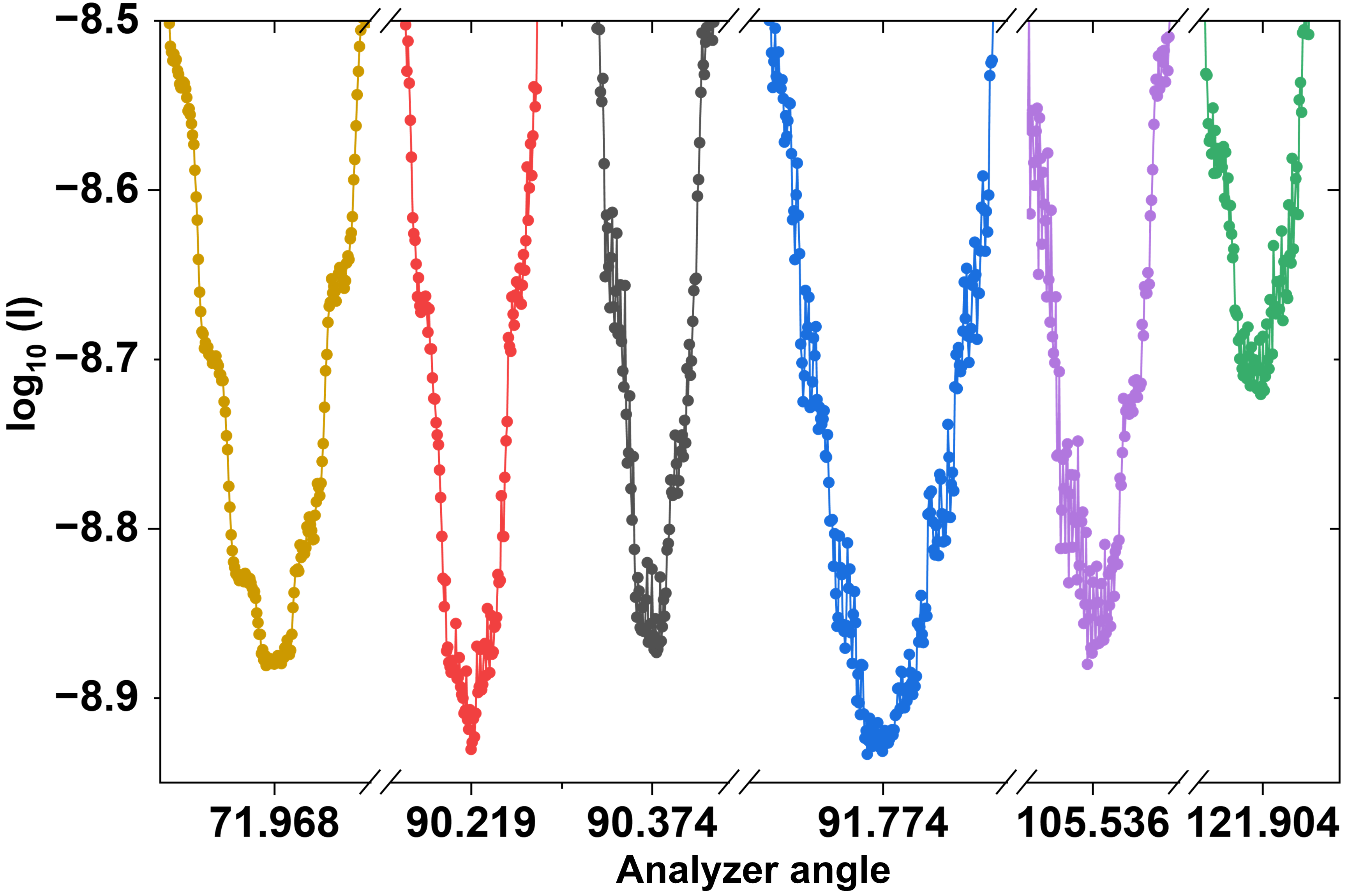}% Here is how to import EPS art
\caption{\label{fig6} Experimentally measured average beam intensity ($I$) in the $20 \times 20$ pixel region centered on the beam axis as a function of analyzer angle, $(\theta_P)$ obtained using the EXE method. The black dots connected by a thin black line represent the laser's reference behavior. The plots for $Q_{||}$, $C_{||}$, $Q_{\perp1}$, $Q_{\perp2}$, and $Q_{\perp3}$ are shown as red, blue, green, purple, and dark yellow dots, respectively, each connected by thin lines of the same color. The different fixed QWP angles $(\theta_Q)$ used to obtain the behavior are given in Table \ref{table 1}.},
\end{figure}

After obtaining the optimized four-lobe structures with a dark-centered cross pattern for both OA orientations (Fig. \ref{fig5}), we select the $20 \times 20$ pixel region centered on the beam axis and extract the average intensity as a function of the analyzer angle. Since the intensity in this region, as a function of $(\theta_A)$ varies by several orders of magnitude (as shown in Fig. 3), we plot the logarithm of the intensity as a function of $(\theta_A)$ (for a fixed $(\theta_Q)$) only in a small angle range for all crystals as shown in Fig. \ref{fig6}. For reference, the log-intensity curve of the input laser beam (without any crystal sample) is also plotted using black dots connected by a thin black line in the figure. The minimum recorded intensity occurs at an analyzer angle of $90.374^\circ$, and the corresponding QWP angle is $-1.303^\circ$, serving as a reference for comparison with crystal samples. The polarization ellipticity of the laser beam was subtracted from subsequent measurements propagating through the crystals to isolate and quantify only the residual EB introduced by the different crystals. The QWP and analyzer angles obtained from the plots (Fig. \ref{fig6}) corresponding to the minimum intensity measured in the dark-central region of (Fig. \ref{fig5}), 
are given in Table \ref{table 1} along with the crystal thickness provided by the manufacturer.

%%%%%%%%%%%%  Fig6  %%%%%%%%%%%%%%%%%
%%%%%%%%%%%%  Fig6  %%%%%%%%%%%%%%%%%

\begin{table}
\caption{\label{table 1}%
The table lists the orientation angles of the QWP and polarizer $(\theta_Q, \theta_P)$ that satisfy the EXE condition for the input laser beam, for different crystal thicknesses and OA orientations, as indicated.}
\begin{ruledtabular}
\begin{tabular}{cccc}
Sample & Thickness (mm) & QWP, $\theta_Q^\circ$ & Analyzer, $\theta_P^\circ$  \\
\hline
Laser & - &  -1.303 & 90.374 \\
$Q_\parallel$ & 1 & -0.895 & 90.219\\
$C_\parallel$ & 1 & -1.739 & 91.774\\
$Q_{\perp1} $ & 1 & -32.762 & 121.904 \\
$Q_{\perp2} $ & 0.5223 & -16.488 & 105.536 \\
$Q_{\perp3} $ & 0.632 & 17.104 & 71.968 \\
\end{tabular}
\end{ruledtabular}
\end{table}

\begin{table*}
\caption{\label{table 2} The polarization ellipticity and ellipse orientation of the transmitted beam through different crystal samples, calculated using Eq. \ref{eq 8} and Eq. \ref{eq 9}, are listed. The residual birefringence, $\delta_{LB}$, for $Q_{\parallel}$ and $C_{\parallel}$, and $\delta_{CB}$, for $Q_{\perp1}$, $Q_{\perp2}$, and $Q_{\perp3}$, of the crystals are calculated from the experimentally obtained $\theta_Q, \theta_P$ angles, using Eq. \ref{eq 21} and Eq. \ref{eq 23}. The refractive index difference $\Delta n$ values calculated from the residual EB of the crystals are compared with literature values. Using Eq. \ref{eq C1} and \ref{eq C2} from Appendix \ref{appendixC}, the $\Delta n$ calculation errors are respectively $Q_{\parallel}$ = $\pm$3.242 $\times 10^{-6}$, $C_{\parallel}$ = $\pm$7.194 $\times 10^{-6}$, and $Q_{\perp1,2,3}$ = $\pm$5.74 $\times 10^{-7}$}. 
\begin{ruledtabular}
\begin{tabular}{cccccc}
 Sample & Ellipticity ($^\circ$) & Ellipse orientation ($^\circ$) & ($\delta_{LB} / \delta_{CB}$) &\begin{tabular}{c}
$|\Delta n|$ \\
$(|n_e-n_o| \text{ or } |n_L-n_R|)$
\end{tabular} & $|\Delta n|$ \\ \hline
 Laser  & -0.9293  & 1.3033 & - & - & - \\
 $Q_\parallel$ & -0.6760 & 0.8950 &  (113.0700 - 0.3558i) & (9.258 - 0.0291i) $\times 10^{-3}$ &9.228 $\times 10^{-3}$ \cite{nagib2003retardation} \\
 $C_\parallel$ & 0.0354 & 1.7386 & (2136.9231 + 0.0830i) & (0.17498 + i 6.79$\times10^{-6}$) & 0.175 \cite{ghosh1999dispersion}  \\
 $Q_{\perp1} $  & -0.8584 & 32.7624 & (1.0975 - 0.0109i) & (8.987 - 0.089i) $\times 10^{-5}$ &  8.15 $\times 10^{-5}$ \cite{song1994color} \\
 $Q_{\perp2} $ & -0.9518 & 16.4878 & (0.5293 - 0.0132i) & (8.297 - 0.207i)  $\times 10^{-5}$ & 8.15 $\times 10^{-5}$ \cite{song1994color} \\
 $Q_{\perp3} $ & -0.9283 & -17.1037 & (0.5507 + 0.0491i) & (8.314 + 0.204i) $\times 10^{-5}$ & 8.15 $\times 10^{-5}$ \cite{song1994color}\\
\end{tabular}
\end{ruledtabular}
\end{table*}

\section{Quantification of Residual Elliptical Birefringence}

The birefringence calculations described in Section \ref{section IV} require the input and transmitted electric fields, ($E^x_{in}, E^y_{in} $) and ($E^x_{out}, E^y_{out} $), respectively. The input electric field, ($E^x_{in}, E^y_{in} $), is first reconstructed from the experimentally determined QWP ($\theta_Q$) and analyzer ($\theta_P$) angles corresponding to the EXE null condition of the incident laser beam and expressed in the amplitude-phase form given by Eq. (\ref{eq 6}). The same EXE procedure is then repeated after introducing the crystal sample into the beam path to reconstruct the transmitted electric field, ($E^x_{out}, E^y_{out} $). These reconstructed electric fields are subsequently substituted into Eqs. (\ref{eq 21}) and (\ref{eq 23}) to calculate the linear ($\delta_{LB}$) and circular ($\delta_{CB}$) birefringence of the crystal, respectively.

It is important to highlight the significance of the proposed and demonstrated EXE method here. Although the laser beam is assumed to be linearly polarized, due to the Brewster-cut window cavity, it is actually elliptically polarized. The calculated ellipticity and ellipse orientation of the laser are given in Table \ref{table 2}. The EXE method's ability to measure such small values is important to extend it to subsequently measure the EB of the crystals. The calculated values of crystal birefringence $\delta_{LR}$ for $Q_\parallel, C_\parallel$ and $\delta_{CB}$ for $Q_{\perp1}, \; Q_{\perp2},\; Q_{\perp3}$ are listed in Table \ref{table 2}. The crystals $Q_\parallel, C_\parallel$ with their OA cut parallel to the surface and with the input field oriented parallel to the OA, are expected to only be linearly birefringent. However, from the values given in Table \ref{table 2}, based on the EXE measurements, we find that the output beam is elliptically polarized due to residual EB introduced by the crystal. Quartz crystals $Q_{\perp1}, \; Q_{\perp2},\; Q_{\perp3}$ with their OA cut perpendicular to the surface and kept along the beam propagation direction are expected to behave as polarization rotators due to circular birefringence. In these crystals, we also observe that the output beam is elliptically polarized, with the plane of polarization rotated away from the input plane. These additional rotations are adjusted by the QWP and the analyzer to different angles to null the beam intensity at the beam center. From these angles, we calculated the residual EB of the crystals. The ellipticity and ellipse orientation of the transmitted beam through the sample crystals calculated using Eq. \ref{eq 8}, Eq. \ref{eq 9}, along with the calculated values of crystal birefringences using Eq. \ref{eq 14} and Eq. \ref{eq 16}, are listed in Table \ref{table 2}. The values calculated using the EXE method are compared with the $\Delta n$ values reported in the literature. Although the real part of the birefringence calculated using the EXE method matches the values reported in the literature, we also observed effects from the imaginary part of the refractive indices, leading to polarization ellipticity in these crystals, which was previously assumed to be non-existent due to measurement limitations. The high sensitivity of the EXE method to measure small changes in the SoEP in the output beam confirms the unexpected EB in these predominantly LB and CB uniaxial crystals.

\section{\label{sec:IX} Summary}
We presented a generalized framework to characterize the residual elliptical birefringence due to weak optical anisotropy in uniaxial crystals by exploiting the intrinsic spin–orbit interaction (SOI) of a paraxial laser beam. Using the vectorial Debye–Richards-Wolf formalism, we simulate the non-uniform spatial polarization structure of an $Ar^+$ laser and confirm its SOI signature experimentally via spatially spin separated four-lobe intensity pattern under EXE conditions. This establishes the sensitivity of the proposed method to measure very small SoEP values in the laser beam and use it to quantify the EB of different uniaxial crystals. We model the transmission of the laser beam through uniaxial crystals in two configurations: with their optic axis parallel or perpendicular to the surface. We derive and use generic transmission matrices that account for the weak EB in otherwise linear or circular birefringent crystals. To quantify the changes in the weak SoEP in the laser beam propagating through the crystals, we extract the average intensity in the $20 \times 20$ pixel region of the dark-centered cross-pattern, corresponding the beam axis as a function of analyzer angle, for fixed $\theta_Q$, achieving a high extinction ratio down to $10^{-8}$ W. This allows us to extract the residual SoEP due to propagation through different crystals and calculate the corresponding birefringence and $\Delta n$ values. Our results demonstrate the effectiveness of using SOI-induced polarization variations in a laser beam to quantify the residual EB of crystals with high accuracy and sensitivity. Our demonstrated ability to measure beam intensity down to the detector's noise floor and use it to calculate the residual EB of the crystals is unique and enables a sensitive, practical tool for anisotropy measurements, extending its applicability to characterize monolayers, 2D materials, ultra-thin films, and metasurfaces. The potential of the EXE method to measure the anisotropy dispersion, improve the measurement sensitivity using detectors with better noise floor and as an alternative to the popular high-accuracy universal polarimeter (HAUP) and ellipsometry methods is undeniable.

\begin{acknowledgments}
The authors acknowledge the Science and Engineering Research Board (SERB), Government of India, for financial support for this work and the Ministry of Social Justice and Empowerment, Government of India, for the research fellowship (UB).
\end{acknowledgments}

\section*{AUTHOR DECLARATIONS}
The authors declare that they have no conflict of interest.

\section*{Data Availability Statement}
Data supporting the findings of this study are available from the corresponding author on a reasonable request.

\appendix

\section{\label{appendixA} Calculation of Linear Birefringence}

For linear birefringence, from Eq. \ref{eq 11}, the output transmitted light field with optic axis azimuth $\phi = 0$ is- 
\begin{align}
    \begin{bmatrix}
         E^x_{out} \\ E^y_{out}
    \end{bmatrix} & = \begin{bmatrix}
        e^{i \frac{\delta}{2}} & 0 \\ 0 & e^{-i \frac{\delta}{2}}
    \end{bmatrix} \begin{bmatrix}
        E^x_{in} \\ E^y_{in}
    \end{bmatrix}   \label{Eq A1}
\end{align} 

 \begin{center}
  So, \hspace{0.03cm}  $ E^x_{out} = e^{i \frac{\delta}{2}} \:  E^x_{in} $ and  $ E^y_{out} = e^{-i \frac{\delta}{2}} \:  E^y_{in} $
\end{center} 
\vspace{-0.5 cm}
\begin{align}
    E^x_{out} \; E^y_{out} = & e^{i \frac{\delta}{2}} \: E^x_{in} \; E^y_{out} \label{eq A2}\\
    E^x_{out} \; E^y_{out} = & e^{-i \frac{\delta}{2}} \: E^y_{in} \; E^x_{out} \label{eq A3}
\end{align}

From Eq. \ref{eq A2} and Eq. \ref{eq A3}, we get - 
\begin{align}
    \cfrac{ E^x_{in} \; E^y_{out}}{E^y_{in} \; E^x_{out}} = e^{-i \delta}  
\end{align}

Considering $\cfrac{ E^x_{in} \; E^y_{out}}{E^y_{in} \; E^x_{out}} = E_L$ , here $E_L$ is a complex number.
\vspace{-0.5 cm}
\begin{align}
    E_L = & \: e^{-i \delta} \notag \\
    ln (E_L) = & i (-\delta + 2 \pi m) \notag\\
    \delta_{LB} = & \: i \: ln(E_L) + 2 \pi m \label{eq A5}  
\end{align}

\section{\label{appendixB} Calculation of Circular Birefringence} 
From Eq. \ref{eq 15}, the output transmitted light field for a circular retarder is -  

\begin{align}
    \begin{bmatrix}
         E^x_{out} \\ E^y_{out}
    \end{bmatrix} = &  \begin{bmatrix}
        \cos \frac{\delta}{2} & - \sin \frac{\delta}{2} \\
       + \sin \frac{\delta}{2} & \cos \frac{\delta}{2}
    \end{bmatrix}  \begin{bmatrix}
        E^x_{in} \\ E^y_{in} 
    \end{bmatrix}  \notag\\
    E^x_{out} = &\cos \frac{\delta}{2} \; E^x_{in} - \sin \frac{\delta}{2} \; E^y_{in} \\
    E^y_{out} = &\sin \frac{\delta}{2} \; E^x_{in} + \cos \frac{\delta}{2} \; E^y_{in} \\ 
    ({E^x_{out}}^2  - {E^y_{out}}^2)  =  & (\cos^2 \frac{\delta}{2} - \sin^2 \frac{\delta}{2}) \; {E^x_{in}}^2 \notag \\
    & - (\cos^2 \frac{\delta}{2} - \sin^2 \frac{\delta}{2}) \; {E^y_{in}}^2 \notag \\
    & - 4 \: \cos \frac{\delta}{2} \: \sin \frac{\delta}{2} \:  E^x_{in} \: E^y_{in} \notag \\ 
    ({E^x_{out}}^2  - {E^y_{out}}^2)  =  &  \: \cos \delta ({E^x_{in}}^2  - {E^y_{in}}^2) - 2 \sin \delta \; E^x_{in} \; E^y_{in} \notag \\
    \biggl( \cfrac{{E^x_{out}}^2  - {E^y_{out}}^2}{{E^x_{in}}^2  - {E^y_{in}}^2} \biggl) = & \: \cos \delta - 2 \sin \delta \; \biggl(\cfrac{ E^x_{in} \; E^y_{in}}{{E^x_{in}}^2  - {E^y_{in}}^2}\biggl)
\end{align}

Considering
\begin{align}
   \biggl( \cfrac{{E^x_{out}}^2  - {E^y_{out}}^2}{{E^x_{in}}^2  - {E^y_{in}}^2} \biggl) =  & E_{C1} \;\; \text{and} \; \; \biggl(\cfrac{ E^x_{in}  E^y_{in}}{{E^x_{in}}^2  - {E^y_{in}}^2}\biggl) = E_{C2} \notag \\ 
     E_{C1} = & \: \cos \delta - 2 \sin \delta \; E_{C2} \notag \\ 
     R \cos (\delta + \alpha ) = & \: \cos \delta - 2 \sin \delta \; E_{C2} \notag \\ 
     R \cos \delta \cos \alpha -   R \sin \delta & \sin \alpha = \: \cos \delta - 2 \sin \delta \; E_{C2} \notag \\
     R \cos \alpha = 1 \hspace{.5 cm} ; & \hspace{.5 cm} R \sin \alpha = 2 E_{C2} \notag \\ 
     R = \pm \sqrt{1+4  E^2_{C2}} \hspace{.5 cm} ; & \hspace{.5 cm} \alpha = \tan^{-1} (2 E_{C2})  \notag \\ 
      E_{C1} =  \pm  \sqrt{1+4  E^2_{C2}} \; & \cos (\delta +  \tan^{-1} (2 E_{C2}) ) \notag \\
      \delta_{CB} = \pm \cos^{-1} \Biggl( \frac{E_{C1}}{\sqrt{1+4  E^2_{C2}}} \Biggl) & - \tan^{-1} (2 E_{C2}) \label{eq B4}
\end{align}

\section{\label{appendixC} Error calculation}
We used a Monte Carlo error analysis to estimate the uncertainties in the reconstructed polarization parameters. The QWP was mounted on a motorized rotation stage (K10CR1/M, Thorlabs, USA) with a specified angular repeatability of, $\sigma_Q= \pm 0.00344^\circ$, while the analyzer was mounted on a NanoRotator stage (NR360S/M, Thorlabs, USA) with a specified angular repeatability of, $\sigma_P= \pm 0.00278^\circ$. These values were used as the angular uncertainties in the Monte Carlo simulations. 

For sampling, 
\begin{align*}
    Q_k &= Q_0 + \sigma_Q N(0,1)\\
    P_k &= P_0 + \sigma_P N(0,1)\\
    \mathrm{Where} \; \sigma_Q & = 0.00344^\circ \mathrm{and} \;  \sigma_P = 0.00278^\circ \\
    \mathrm{and} \; k & = 1,....,N  \mathrm{with} \; N = 5000
\end{align*}

Sampling parameters:
\begin{itemize}
    \item $Q_0$: measured QWP angle and $P_0$: measured analyzer angle
    \item $Q_k$: QWP and $P_k$: analyzer angle in the $k$-th Monte Carlo realization.
    \item$ N(0,1)$: standard normal random variable with mean 0 and standard deviation 1. 
    \item $N$: total number of Monte Carlo realizations ($N=5000$). 
    \item $k$: realization index ($k=1,2,…,N$).
\end{itemize}

Using Eq. \ref{eq 5}, the amplitude, a, and phase, $\Delta$, are reconstructed from Eq. \ref{eq 6}. In the following discussion, the phase will be denoted by ($\zeta$) to distinguish it from the uncertainty notation used in the error analysis.
For each realization,
\begin{equation*}
    (a_k,\zeta_k)=f(Q_k,P_k )
\end{equation*} 
where $f$ is the polarization-state reconstruction algorithm. 

\noindent Now, using the Monte Carlo error analysis, 

\noindent Amplitude uncertainty,  $\Delta A=std(a_k)$ ; 
here, $std(.)$ is the standard deviation of all Monte Carlo realizations.

\noindent Phase uncertainty (after phase unwrapping)
\begin{center}
    $\Delta \zeta=std(\zeta_k)$
\end{center}
 
\noindent Ellipticity, \vspace{-0.5 cm}
\begin{align*}
  & \epsilon_k  = \cfrac{1}{2} \sin^{-1} [\sin (2 a_k) \sin(\zeta_k)] \\
    & \Delta \epsilon  = std(\epsilon_k) \\
\end{align*}

\vspace{-0.5 cm}
\noindent Ellipse orientation, \vspace{-0.3 cm}
\begin{align*}    
    & \phi_k  = \cfrac{1}{2} \tan^{-1} [\tan (2 a_k) \cos(\zeta_k)] \\
    & \Delta \phi  = std(\phi_k)\\
\end{align*}

\vspace{-0.5 cm}
\noindent Linear retardance, \vspace{-0.3 cm}
\begin{align*}
   & \delta_{LB,k}  = \zeta_{0,k} - \zeta_{i,k}\\
    & \Delta \delta_{LB} = std(\delta_{LB,k}) 
\end{align*}

\noindent Linear birefringence, \vspace{-0.3 cm}
\begin{align}
    \Delta n_{LB,k} & = \cfrac{\lambda}{2 \pi d} \:\delta_{LB,k}  \notag \\
    (\sigma_{\Delta n})_{LB} & = std(\Delta n_{LB,k}) \label{eq C1}
\end{align}

\noindent Circular retardance,
\begin{align*}
  & \hspace{1cm} U_k  = \tan(2a_{i,k}) \cos(\zeta_{i,k})\\
  & \hspace{1cm} D_k  = \sqrt{1+U_k^2 } \\
  & \hspace{1cm} \delta_{CB,k}  = \cos^{-1} \Big[\cfrac{\cos(2a_{o,k})}{\cos(2a_{i,k})D_k} \Big] - \tan^{-1}(U_k) \\
  & \hspace{1cm} \Delta \delta_{CB}  = std( \delta_{CB,k})
\end{align*}

\noindent Circular birefringence, \vspace{-0.3 cm}
\begin{align}
   & \Delta n_{CB,k}  = \cfrac{\lambda}{2 \pi d} \:\delta_{CB,k} \notag \\
   & (\sigma_{\Delta n})_{CB}  = std(\Delta n_{CB,k}) \label{eq C2}
\end{align}

Using the above-mentioned method, errors in $\Delta n$ are calculated for $Q_\parallel$, $C_\parallel$, and $Q_{\perp1,2,3}$.

% The \nocite command causes all entries in a bibliography to be printed out
% whether or not they are actually referenced in the text. This is appropriate
% for the sample file to show the different styles of references, but authors
% most likely will not want to use it.
%\nocite{*}
\newpage
\bibliography{apssamp}% Produces the bibliography via BibTeX.

@PREAMBLE{
 "\providecommand{\noopsort}[1]{}" 
 # "\providecommand{\singleletter}[1]{#1}%" 
}

@article{malykin2016vl,
  title={VL Ginzburg’s helical elliptically polarized modes and their application},
  author={Malykin, Grigorii B},
  journal={Physics-Uspekhi},
  volume={59},
  number={12},
  pages={1245},
  year={2016},
  publisher={IOP Publishing}
}

@incollection{ramachandran1961crystal,
  title={Crystal optics},
  author={Ramachandran, GN and Ramaseshan, S},
  booktitle={Kristalloptik{\textperiodcentered} Beugung/Crystal Optics{\textperiodcentered} Diffraction},
  pages={1--217},
  year={1961},
  publisher={Springer}
}

@article{nye1985physical,
  title={Physical properties of crystals, Clarendon},
  author={Nye, JF},
  journal={Oxford},
  volume={19762},
  pages={3},
  year={1985}
}

@article{kobayashi1983new,
  title={A new optical method and apparatusHAUP'for measuring simultaneously optical activity and birefringence of crystals. I. Principles and construction},
  author={Kobayashi, J and Uesu, Y},
  journal={Applied Crystallography},
  volume={16},
  number={2},
  pages={204--211},
  year={1983},
  publisher={International Union of Crystallography}
}

@article{chou1997effect,
  title={Effect of elliptical birefringence on the measurement of the phase retardation of a quartz wave plate by an optical heterodyne polarimeter},
  author={Chou, Chien and Huang, Yeu-Chuen and Chang, Ming},
  journal={Journal of the Optical Society of America A},
  volume={14},
  number={6},
  pages={1367--1372},
  year={1997},
  publisher={Optical Society of America}
}

@article{arteaga2009determination,
  title={Determination of the components of the gyration tensor of quartz by oblique incidence transmission two-modulator generalized ellipsometry},
  author={Arteaga, Oriol and Canillas, Adolf and Jellison Jr, Gerald E},
  journal={Applied optics},
  volume={48},
  number={28},
  pages={5307--5317},
  year={2009},
  publisher={Optical Society of America}
}

@article{martin2021chiroptical,
  title={Chiroptical anisotropy of crystals and molecules},
  author={Martin, Alexander T and Nichols, Shane M and Murphy, Veronica L and Kahr, Bart},
  journal={Chemical Communications},
  volume={57},
  number={66},
  pages={8107--8120},
  year={2021},
  publisher={Royal Society of Chemistry}
}

@article{yang2023simultaneous,
  title={Simultaneous measurement of polarization rotation angle and ellipticity at the quantum noise limit},
  author={Yang, Peng and Xie, Boya and Feng, Sheng},
  journal={Journal of the Optical Society of America B},
  volume={40},
  number={11},
  pages={2900--2905},
  year={2023},
  publisher={Optica Publishing Group}
}

@article{ciattoni2002paraxial,
  title={Paraxial propagation along the optical axis of a uniaxial medium},
  author={Ciattoni, Alessandro and Cincotti, Gabriella and Provenziani, Damiano and Palma, Claudio},
  journal={Physical Review E},
  volume={66},
  number={3},
  pages={036614},
  year={2002},
  publisher={APS}
}

@article{ciattoni2003circularly,
  title={Circularly polarized beams and vortex generation in uniaxial media},
  author={Ciattoni, Alessandro and Cincotti, Gabriella and Palma, Claudio},
  journal={Journal of the Optical Society of America A},
  volume={20},
  number={1},
  pages={163--171},
  year={2003},
  publisher={Optical Society of America}
}

@article{volyar2002vector,
  title={Vector singularities of Gaussian beams in uniaxial crystals: optical vortex generation},
  author={Volyar, AV and Fadeeva, TA and Egorov, Yu A},
  journal={Technical physics letters},
  volume={28},
  number={11},
  pages={958--961},
  year={2002},
  publisher={Springer}
}

@article{allen2003optical,
  title={Optical angular momentum. IOP Publishing Ltd and individual contributors},
  author={Allen, L and Barnett, SM and Padgett, MJ},
  journal= {IOP Publishing Ltd and individual contributors},
  year={2003}
}

@article{simpson1997mechanical,
  title={Mechanical equivalence of spin and orbital angular momentum of light: an optical spanner},
  author={Simpson, NB and Dholakia, K and Allen, L and Padgett, MJ},
  journal={Optics letters},
  volume={22},
  number={1},
  pages={52--54},
  year={1997},
  publisher={Optical Society of America}
}

@article{liberman1992spin,
  title={Spin-orbit interaction of a photon in an inhomogeneous medium},
  author={Liberman, VS and Zel’dovich, B Ya},
  journal={Physical Review A},
  volume={46},
  number={8},
  pages={5199},
  year={1992},
  publisher={APS}
}

@article{marrucci2006optical,
  title={Optical spin-to-orbital angular momentum conversion in inhomogeneous anisotropic media},
  author={Marrucci, Lorenzo and Manzo, Carlo and Paparo, Domenico},
  journal={Physical review letters},
  volume={96},
  number={16},
  pages={163905},
  year={2006},
  publisher={APS}
}

@article{lerman2008generation,
  title={Generation of a radially polarized light beam using space-variant subwavelength gratings at 1064 nm},
  author={Lerman, Gilad M and Levy, Uriel},
  journal={Optics Letters},
  volume={33},
  number={23},
  pages={2782--2784},
  year={2008},
  publisher={Optical Society of America}
}

@article{levy2019mathematics,
  title={Mathematics of vectorial Gaussian beams},
  author={Levy, Uri and Silberberg, Yaron and Davidson, Nir},
  journal={Advances in Optics and Photonics},
  volume={11},
  number={4},
  pages={828--891},
  year={2019},
  publisher={Optical Society of America}
}

@article{simon1987cross,
  title={Cross polarization in laser beams},
  author={Simon, R and Sudarshan, ECG and Mukunda, N},
  journal={Applied optics},
  volume={26},
  number={9},
  pages={1589--1593},
  year={1987},
  publisher={Optical Society of America}
}

@article{fainman1984polarization,
  title={Polarization of nonplanar wave fronts},
  author={Fainman, Yeshaiahu and Shamir, Joseph},
  journal={Applied optics},
  volume={23},
  number={18},
  pages={3188--3195},
  year={1984},
  publisher={Optical Society of America}
}

@article{erikson1994polarization,
  title={Polarization properties of Maxwell-Gaussian laser beams},
  author={Erikson, WL and Singh, Surendra},
  journal={Physical Review E},
  volume={49},
  number={6},
  pages={5778},
  year={1994},
  publisher={APS}
}

@article{sokolov2002polarization,
  title={Polarization of spherical waves},
  author={Sokolov, AL},
  journal={Optics and Spectroscopy},
  volume={92},
  number={6},
  pages={936--942},
  year={2002},
  publisher={Springer}
}

@article{baishya2022dark,
  title={Dark-field spin Hall effect of light},
  author={Baishya, Upasana and Kumar, Nitish and Viswanathan, Nirmal K},
  journal={Optics Letters},
  volume={47},
  number={17},
  pages={4479--4482},
  year={2022},
  publisher={Optica Publishing Group}
}

@article{kumar2025spin,
  title={Spin--orbit conversion-enabled generation of cylindrical vector beams},
  author={Kumar, Nitish and Viswanathan, Nirmal K},
  journal={Optics Communications},
  pages={132089},
  year={2025},
  publisher={Elsevier}
}

@article{benelajla2021physical,
  title={Physical origins of extreme cross-polarization extinction in confocal microscopy},
  author={Benelajla, Meryem and Kammann, Elena and Urbaszek, Bernhard and Karrai, Khaled},
  journal={Physical Review X},
  volume={11},
  number={2},
  pages={021007},
  year={2021},
  publisher={APS}
}

@article{steindl2023cross,
  title={Cross-polarization-extinction enhancement and spin-orbit coupling of light for quantum-dot cavity quantum electrodynamics spectroscopy},
  author={Steindl, P and Frey, JA and Norman, J and Bowers, JE and Bouwmeester, D and L{\"o}ffler, W},
  journal={Physical Review Applied},
  volume={19},
  number={6},
  pages={064082},
  year={2023},
  publisher={APS}
}

@misc{azzam1978ellipsometry,
  title={Ellipsometry and polarized light},
  author={Azzam, Rasheed MA and Bashara, Nicholas Mitchell and Ballard, Stanley S},
  year={1978},
  publisher={American Institute of Physics}
}

@book{born2013principles,
  title={Principles of optics: electromagnetic theory of propagation, interference and diffraction of light},
  author={Born, Max and Wolf, Emil},
  year={2013},
  publisher={Elsevier}
}

@article{nagib2003retardation,
  title={Retardation characteristics and birefringence of a multiple-order crystalline quartz plate},
  author={Nagib, NN and Khodier, SA and Sidki, HM},
  journal={Optics \& Laser Technology},
  volume={35},
  number={2},
  pages={99--103},
  year={2003},
  publisher={Elsevier}
}

@article{ghosh1999dispersion,
  title={Dispersion-equation coefficients for the refractive index and birefringence of calcite and quartz crystals},
  author={Ghosh, Gorachand},
  journal={Optics communications},
  volume={163},
  number={1-3},
  pages={95--102},
  year={1999},
  publisher={Elsevier}
}

@article{song1994color,
  title={Color separation of argon laser light with the effect of dispersion of optical activity},
  author={Song, Feijun and Yu, Lei and Yang, Yanfeng and Tan, Hong},
  journal={Applied optics},
  volume={33},
  number={24},
  pages={5513--5517},
  year={1994},
  publisher={Optical Society of America}
}

@article{hong2011background,
  title={Background-free detection of single 5 nm nanoparticles through interferometric cross-polarization microscopy},
  author={Hong, Xin and van Dijk, Erik MPH and Hall, Simon R and G{\"o}tte, J{\"o}rg B and van Hulst, Niek F and Gersen, Henkjan},
  journal={Nano letters},
  volume={11},
  number={2},
  pages={541--547},
  year={2011},
  publisher={ACS Publications}
}

@article{baishya2023measurement,
  title={Measurement of surface chirality at near-normal incidence},
  author={Baishya, Upasana and Viswanathan, Nirmal K},
  journal={Applied Physics Letters},
  volume={122},
  number={26},
  year={2023},
  publisher={AIP Publishing}
}

@article{richards1959electromagnetic,
  title={Electromagnetic diffraction in optical systems, II. Structure of the image field in an aplanatic system},
  author={Richards, Bernard and Wolf, Emil},
  journal={Proceedings of the Royal Society of London. Series A. Mathematical and Physical Sciences},
  volume={253},
  number={1274},
  pages={358--379},
  year={1959},
  publisher={The Royal Society London}
}

\end{document}